\documentclass[14pt]{article}
\usepackage[dvips]{graphicx}
\date{}
\title{
 PRODUCTION OF LIGHT GOLDSTONE PARTICLES ON PHOTON COLLIDERS\\
}
\author{S.I.Polityko \\
\it {Irkutsk State University}\\}
\begin{document}

\maketitle
\begin{abstract}
         By realizing the project of intensive
 $\gamma $ beams with large energy (necessary for Photon  Colliders)
 an essential flux
of light Goldstone particles (LGP - axions, arions, familons, majorons)
can be generated. The probability of LGP production is calculated
for different densities of laser photons in the conversion region
(with absorption of one or several laser photons).
The method for observation of LGP via its absorption in the matter
is also presented.
\end{abstract}

\section {Introduction}
        The existence of Light Goldstone Particles (LGP),
usually pseudoscalar (axion, arion, majoron, familon), is very
attractive in many theoretical schemes (for recent review see
e.g. \cite{Ans,Sik}). Such particles are still elusive. So, the new
method of their possible discovery looks attractive. In this respect,
the program of construction of Photon Colliders based on future
Linear Colliders provides new opportunity.

        To describe this opportunity we remind necessary features
of obtaining  Photon Colliders. The intense photon beams for Photon
Collider should be obtained by backscattering of electrons prepared
for the Linear Collider on the intense laser light
 \cite{GKST,Gin,Tel,Brink}. It is expected
 to obtain here the conversion coefficient $k$ ( the ratio of obtained high
energy photons to the number of initial electrons $N_e$) close to 1.
The main features of conversion are described by quantity $x$ which determined
via initial electron beam energy $E$ and laser photon energy $\omega_0$
as
\begin{equation}
\label{1}
x={4E\omega_0\over m_e^2}\,.
\end{equation}
The best quality of photon spectra is obtained at highest values of $x$,
which are limited from above by a value about $x_0=2(1+\sqrt{2})\approx 4,8$.
The conversion coefficient $k=N_\gamma \sigma_C/S$, where $N_\gamma$ is the
number of photons in the laser bunch, $S$ -- its cross section in the
conversion point and $\sigma_C$ is the standard cross section of Compton
scattering at c.m. energy $W_0=m_e\sqrt{x+1}$.

From this it
follows that  the conversion region can be considered as
$e\gamma_0$ collider ($\gamma_0$ is the laser photon) with huge luminosity
${\cal L}$ but small c.m.s. energy $W_0$:
\begin{equation}
\label{2}
{\cal L}= f{N_eN_\gamma\over S}=f{kN_e\over \sigma_C}\approx
10^{45}\div 10^{46} \rm{ cm}^{-2}\rm{year}^{-1}\,,\quad
W_0=m_e\sqrt{x+1}\approx 1.2\rm{ MeV}.
\end{equation}
Here $N_e$ is the number of electrons in bunch and $f$ is the repetition rate.

Therefore, this conversion region presents unique place for the study of
LGP production if its mass is less than $W_0-m_e$. This opportunity was
studied in Refs.~\cite{Pol1,Brod,Pol2,GKP}. In this paper we discuss new
potential given by effects of high laser photon density in the
conversion region. This high density makes strongly probable the processes
\begin{equation}
\label{3}
e+n\gamma_0=e+X.
\end{equation}
with simultaneous absorption of $n>1$ laser photon\footnote {
 In the second mechanism of production
 $n\gamma_0+\gamma \rightarrow X$ (where $\gamma $  is the high energy
 photon)  the number of produced LGP's is essentially less than that
discussed in the text the corresponding cross section contains additional
 factor $m_X^2/m_e^2$}.
 Evidently, in such a
process the permissible region of LGP mass is expanded to values $\sim
m_e(\sqrt{nx+1}-1)$.

To this goal we discuss briefly modern status of different LGP (section 2).
Next we describe the method to calculate probabilities of processes under
interest in the field of intense laser light (section 3). Last we discuss
the proposed registration method (section 4).

\section{ Various LGP's. Modern status.}

 The Effective Lagrangian describing an interaction of any LGP $X_i$
with electrons and photons has the form
\begin{equation}
\label{4}
{\cal L}= g_{iee}\bar e \gamma _5 eX_i+C_{i\gamma \gamma}{\alpha \over m_e}
 X_iF\tilde F
\end{equation}

\subsection {The "invisible" axion}

The axion is a light pseudogoldstone boson, proposed
 for the solution of the CP-violation problem in strong interactions
 \cite{SW,FW}. The existence of this standard axion is practically
forbidden by modern data (see e.g \cite{GGR}). However,
 the idea of natural explanation of CP symmetry is attractive, and the
theory of the  standard
 axion was modified to make its interaction with matter weaker and to
 make it lighter.

The axion model with two Higgs doublets is characterized by the scale of
 breaking  of the $U(1)$-symmetry $f_{pq} \approx 246$ GeV and the mass
$m_a \geq 150$ KeV, connected by one parameter: the ratio of the vacuum
expectation value (VEV) of Higgs doublets.

 The addition of new Higgs multiplets breaks relation between $m_{a}$ and
standard  $f_{pq}$ value. It happens that
 $250$ GeV $\leq f_{pq} \leq 10^{19}$ GeV, the mass $m_{a}$, just as
the coupling with matter can become sufficiently smaller than in the basic
theory \cite{Ans}. For example, one can introduce the
additional scalar field
 ($SU(2)\times U(1)$ singlet) with arbitrary large VEV.
 The obtained "invisible" axion $X_a$ is known as
Dine-Fischler-Srednicki-Zhitnitsky  axion (DFSZ) \cite{Z,DFS}.

The couplings $g_{aee}$ and $C_{a\gamma \gamma}$ in eq. (\ref{4})
are given by:
\begin{equation}
 g_{aee}={m_am_e\over f_\pi m_\pi}{1+z\over N\sqrt{z}}\cos^2\beta
\approx 2.1{m_am_e\over f_\pi m_\pi}\cos^2\beta,
\label{5}
\end{equation}
\begin{equation}
 C_{a\gamma \gamma}={m_am_e\over 8\pi f_\pi m_\pi}2\sqrt{z}\approx
0.06{m_am_e\over  f_\pi m_\pi},
\label{6}
\end{equation}
where $f_\pi=94$ MeV is the pion decay constant, $z=m_u/m_d=0.568$
 is the ratio of the quark masses, $N$ is the number of generations,
 $\tan\beta$ is the ratio of the VEV's of two Higgs fields.
 Respectively, the mass of axion is given by
\begin{equation}
 m_a=m_\pi{f_\pi\over f_a\sin 2\beta}{2N\sqrt{z} \over 1+z}\approx
0.96m_\pi{Nf_\pi\over f_a\sin 2\beta}.
\label{7}
 \end{equation}
Another form of "invisible" axion was proposed in
 Refs.~\cite{K,SVZ},
 now it is known as hadronic or Kim-Shifman-Vainstein-Zakharov
 axion (KSVZ).
 This axion does not interact with leptons at the tree level.
 Therefore, its couplings  to leptons and photons
 are two orders less compared to DFSZ axion.

The archion model \cite{Ber} contains  the global symmetry $U(1)$,
its  spontaneous breaking results in Goldstone boson,
 which has both diagonal and nondiagonal flavor interaction with fermions.
 But unlike the axion the archion has no interaction with
 photons and it is like a hadronic axion with strongly suppressed lepton
 interaction.

Modern experimental limitation for electron-axion coupling
gives $g_{aee}<3\cdot 10^{-8}$ \cite{PDG} (In this review limitation for
pseudovector coupling $G_{aee}=g_{aee}/(2m_e)$ is presented.)

 The astrophysical constraints on the VEV, mass and coupling of "invisible"
axion   \cite{Kim} are very restrictive:
$$
10^9\rm{ GeV}<f_a<10^{12}\rm { GeV}\Rightarrow
 0.5\cdot 10^{-12} < g_{aee} < 0.5\cdot 10^{-9},
$$
$$
 \quad 0.6\cdot 10^{-5}eV
< m_a < 0.6\cdot 10^{-2}eV.
$$

 \subsection {The Arion $X_\alpha$}

         The arion is a neutral, strictly massless, stable pseudoscalar
 boson with even charge parity,  interacting with fermions \cite{Ans}.
 The interaction of arion with lepton is described by an Effective Lagrangian
(\ref{4}) with:
\begin{equation}
\label{8}
 g_{\alpha ee}=\tan\beta{m_e\over v},\quad
 v=(G_F\sqrt{2})^{-1/2}=246\quad GeV.
 \end{equation}
 Here  $X_\alpha$ is the arion field, $\tan\beta$ is
 the ratio of different VEVs, modern estimations in Higgs boson physics
are in favor of $\tan\beta=1\div 40$. However, the astrophysical
data are in favor of very small $\tan\beta$ (which are not forbidden yet
in the standard two doublet Higgs models).
 Weakly interacting with a matter the arions can be emitted from stars.
 The arion emission leads to fast loss of energy from the stars.
 The demand of the condition by which the arion luminosity of the Sun
 should not exceed the photon luminosity leads to $\tan\beta<10^{-3}$.
 A more strong constraint appears from the evolution of red giants is
 $\tan\beta<10^{-6}$.

 It gives weak coupling with leptons
$g_{aee} <2\cdot 10^{-9}- 2\cdot 10^{-6}$.

 \subsection { The Majoron $X_M$}

     A spontaneously broken global symmetry of lepton number will lead to
 massive Majorana neutrinos and a Nambu-Goldstone boson, named the majoron.
 This
 can be accomplished by extending the Standard Model with an additional
 gauge-single Higgs field \cite{RM}, or $SU(2)-$triplet Higgs multiplet
\cite{GR}. The respective Goldstone bosons are the Chikashige-Mohapatra-
Peccei (CMP) Majoron and the Gelmini-Roncadelli  (GR) Majoron.

        CMP model requires the addition of a gauge-singlet Higgs field and a
right-handed heavy neutrino. The effective Majoron -electron interaction
induced at one-loop level is given by
 \begin{equation}
{\cal L} = {G_F\over 16\pi ^2}m_e m_\nu \bar e\gamma_5 e X_M
\label{9}
\end{equation}
    From the current upper bound $m_{\nu_e} < 10\, eV$ it is easily seen
that the coupling of the CMP Majoron to matter is extremely small
($g_{Mee} < 10^{-17} $).

        An $SU(2)-$triplet Higgs multiplet is introduced in GR Majoron
model. The wave function of the GR Majoron is primarily the phase
field of the neutral component and has a small admixture of the Higgs
doublet with the mixing angle $2v_T/v_D$ ($v_T$ and $v_D$ being the VEV's
of the Higgs triplet and doublet, respectively).
As a consequence, the GR Majoron has a tree-level coupling to electrons
\begin{equation}
{\cal L} = 2\sqrt{2}G_Fv_T m_e \bar e\gamma_5 e X_M
\label{10}
\end{equation}

The astrophysical constraint from the consideration of majoron emission
 rates of the neutron-star core is $v_T< 2 KeV$.

 Thus, we have the following constraint on the coupling with an electron:
 $g_{M ee}<3.4\cdot 10^{-14}$.

 \subsection {The Familon $X_F$}

    The familon is the Goldstone boson associated with the spontaneous
 breaking of a global family symmetry (horizontal symmetry) between
 the generation of the quarks and leptons \cite{Cheng}.
 Since the breaking of the horizontal group can takes place at small
 distance $1/F$ only,
 the familon,  similar to the invisible axion, interacts weakly with the
 matter and its mass should be small.

 The Effective Lagrangian for the interaction of a familon with a
 lepton at low energy has form (\ref{4}) with
 $g_{Fee}=2m_e/F$.

The absence of decay $K^+\rightarrow \pi^+X_F$ results in
limitation $g_{Fee}< 0.8\cdot 10^{-14}$ ($F > 1.3\cdot
 10^{11}$GeV) \cite{PDG}.  The astrophysical constraint
here is less restrictive, it gives $g_{Fee}=2m_e/F<1.4\cdot 10^{-13}$
($F>7\cdot 10^{9}$ GeV).

\section{ The calculations of probability and the number of events.}

      We consider the laser wave with circular
 polarization (which is  the best for conversion \cite{GKST}).
 Its vector potential has  form
  $A_{\mu } = a_{1\mu }\cos \varphi  + a_{2\mu } \sin \varphi,\;
  \varphi =kx$, where $k_{\mu}$ is the momentum of laser photon,
  ($ka_{1}=ka_{2}=a_{1}a_{2}=0,
  a_{1}^{2}=a_{2}^{2}=a^{2}$).

  The matrix element of the LGP production by an electron
in external field can be written in the form
\begin{equation}
     M_{fi}=ig_{Xee} \int d^{4}x \bar \Psi _{e}(x)
     \gamma _{5} \Psi _{e}{\Phi}(x),
\label{11}
\end{equation}
     \par
     \noindent
      where  $\Psi _{e}(x)$ and $\bar \Psi _{e}(x)$ are the exact
 solution
 of the Dirac equation for an electron in the field of a circularly
 polarized wave:
\begin{equation}
     \Psi _{e}(x)=\left(1+{ek\hat A \over 2kp}\right)u_{p}
\exp\left(ie{a_{1}p \over kp} \sin\varphi -ie{a_{2}p \over kp} \cos\varphi
  +iqx\right), \quad \phi _{X}={e^{-ip_{X}x} \over \sqrt{2\epsilon _{X}}}.
\label{12}
\end{equation}
     \par
     \noindent
    Here $\epsilon _{X},p_{X}$ is the energy and the momentum of LGP,
   $p$ and $q$ are the momentum and "quasimomentum"  of an electron
     $$
     q_{\mu } =p_{\mu }+\xi ^{2}{m_{e}^{2} \over 2kp}k_{\mu },
     \quad \xi ^{2}={e^{2}a^{2} \over m_{e}^{2}}.
     $$
Note that
$$
m^2_*=q_\mu q_\mu=m_e^2(1+\xi^2)\,.
$$
This difference from $m_e^2$ is essential to find the production threshold.

    Using standard techniques (see \cite{Ritus,LD}), we calculated
 the probability of LGP production by a nonpolarized electron
 \begin{equation}
 dW={g_{X ee}^2 m_e\over 64\pi }\sum W_{n}
                     {du\over (1+u)^2}
\label{13}
\end{equation}
 $$
  W_{n}= {1\over \sqrt{nx+1+\xi ^2}}
 \left (-2\eta ^2 J_n^2 (z)+\xi ^2 {u^2\over 1+u}[J_{n+1}^2(z)+J_{n-1}^2(z)
                    -2J_n^2(z)] \right ),
 $$
 \par
 \noindent
 where $J_{n}(z)$ is the Bessel function of $n$-th order,
 $u=(kp_a)/(kp^{'} )$, $p^{'}$ is the momentum of a scattering electron,
 $$
 \eta ={m_X\over m_e},\, z={2\xi \over x}\sqrt{nx+1+\xi ^2}\sqrt{{
 (u_+-u)(u-u_-)\over (1+u_+)(1+u_-)}},\,u_-<u<u_+,
 $$
 \par
 $$
 u_{\pm}={nx+\eta ^2 \pm \sqrt{(\eta ^2-nx)^2-4\eta ^2(1+\xi ^2)}\over
 2+nx+2\xi ^2 -\eta ^2 \mp  \sqrt{(\eta ^2-nx)^2-4\eta ^2(1+\xi ^2)}}.
 $$
 \par

 In formula (\ref{13}) the term with given $n$ describes the LGP
 production by an electron via absorption from electromagnetic wave $n$
 of laser photons simultaneously (\ref{3}),
 $n_{th}$ is the minimal number of photons for the Goldstone production
 with the mass $m_X$
 \begin{equation}
 n_{th}={1 \over x}(\eta ^2 +2\eta \sqrt{1+\xi ^2})
\label{14}
\end{equation}
    Integrating the expression (\ref{13}) we get the total  probability of the LGP
production (in this case the term ``cross section'' cannot be used):
 $$
 {g_{X ee}^2 m_e\over 64\pi }f(x,\xi,\eta).
 $$
 In Figure 1 the dependence of function $f(x,\xi,\eta)$ shown at some
 values $\xi$.

\begin{figure}
\centering\includegraphics[bb=170 160 430 650, scale=0.5]{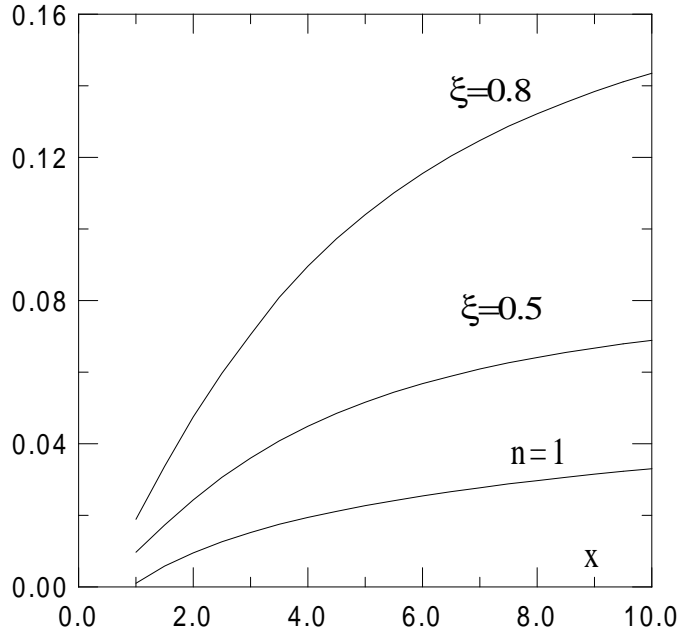}
\caption{The dependence $f(x,\xi)$ on $x$ at some values $\xi$
 and $\eta =0$, curve n=1 corresponds to single photon absorption}
\end{figure}

At $ \xi ^2 \ll 1$, expanding the Bessel function in series, we get from
 (\ref{13})
 \begin{equation}
 dW_n={g_{X ee}^2m_e(\xi /x)^{2n} \over 64\pi n!\sqrt{nx+1+\xi ^2}}
 D {du\over (1+u)^2}
 \label{15}
\end{equation}
 $$
 D=
 -2\eta ^2((1+\xi ^2)(u_+-u)(u-u_-))^n+{x^2n^2u^2\over 1+u}
 ((1+\xi ^2)(u_+-u)(u-u_-))^{n-1}
 $$
   For $n=1$, integrating (\ref{15}) by $u$, we get the production cross-section
   calculated in Ref.~\cite{Pol1}:
 $$
 \sigma ={1\over 2}{\alpha g_{X ee}^2\over m_e^2}{1\over x}
 \Biggl \{ \Biggl(1-2{\eta ^2\over x}+{2\eta^2(\eta^2-2)\over x^2}
 \Biggr)\ln
 \Biggl({4(x+1)\over 2+x-\eta^2+\sqrt{(x-\eta^2)^2-4\eta^2}}\Biggr)\Biggr.
 $$
 $$
 \Biggl.
 + \sqrt{(x-\eta^2)^2-4\eta^2}\Biggl( 1-{7\over 2}\eta^2 +{3\over 2}x
 -8{\eta^2 \over x}-7{\eta^2\over x^2}\Biggr)\Biggr\}
 $$
        At $x=5$  this cross-section is:
 $$
 \sigma \approx g_{X ee}^2\cdot 5.4\cdot 10^{-24} cm^2
 $$

{\large \bf Numerical estimations}. Here we use the physical parameters of
 the laser conversion region which are chosen in accordance with the projects
 of Photon Colliders  \cite{Gin,Tel,Brink}.
         In the conversion scheme, it has been proposed to use a
 laser with photon energy $\omega _{0}=1.17$ eV.
 The length of conversion region characterized by a high density
 of laser photons is  $l \sim 1.5$ nm, which value is close to the length
 of the electron bunch.
 The invariant mass of the $\gamma _{0}e$--system for the energy of
 electron
 $E_e=250$ GeV is comparable to the electron mass
 $W_0 \approx 1.21$ eV ($x=4.5$).
 It is convenient to express $\xi ^2$ in terms of the energy $A$,
 the duration $\tau $ and radius $a_{\gamma }$ of the laser flash
 in the interaction point
 $$
 \xi ^{2} = {A \over A_{*}},\quad {\rm where}\quad A_{*} =
 {\tau c \over 4} \cdot \left ({ m_{e} \omega _{0} a_{\gamma} c \over
 e \hbar } \right ) ^{2}.
 $$
 \par
 \noindent
 For the values $\pi a_{\gamma }^{2} \approx 10^{-5} \; cm^2$ we have
 $A_{*} = 100 $ J and at the energy of laser flash $A = 25$ J
 ${\xi}^{2} =0.25$.
 The value ${\xi}^{2}$ can reach 0.6. The number of produced LGP's is
\begin{equation}
 N_X={N_e\tau \over 2} \sum_{n>n_{th}} \int _{u_-}^{u_+}dW_n
\label{16}
\end{equation}
   The LGP energies are distributed in the interval
\begin{equation}
 {\eta ^2\over (x+\eta ^2)R}E_e<\epsilon _X<E_e {x+\eta ^2\over x+1}R,
\rm{ where}\, R={1\over 2}\left (1+\sqrt{1-{4\eta ^2(x+1)\over (x+\eta ^2
 )^2}} \right )
\label{17}
\end{equation}
 For $E_e=250$ GeV at  $m_X=10$ KeV this corresponds to the interval
  $$20\, MeV<\epsilon _X< 208\,GeV$$.
 Since the effective mass of the $\gamma _0 e$--system is not large
 the characteristic emission angles of Goldstones relative to the
 direction of the electron bunch are $\leq m_e/E_e \approx 10^{-5}$.
 Therefore the angular spread of Goldstones is defined by the angular
 spread of electrons in the beam  ($\approx 10^{-4}$).

    The numerical calculation of the number of LGP's
 and the used couplings are shown in the table. It is seen that
a large number of axions or arions will generate from the discussed
conversion.

 \centerline{\large Table}

 \begin{center}
 \begin{tabular}{|c|c|c|c|}\hline
  & $g_{Xee}$ & Cross-section $\sigma (cm^2)$ & The number of LGP per
 year\\
 \hline  Standard axion & $2\cdot 10^{-6}$ &$2.2\cdot 10^{-35}$ &$7\cdot
 10^{10}$ \\
 \hline  "Invisible" axion &$3\cdot 10^{-8}$ &$4.9\cdot 10^{-39}$
 &$1.5\cdot 10^7$ \\
 \hline  Arion & $2\cdot 10^{-6}$ & $2\cdot 10^{-35}$ &$7\cdot 10^{10}$ \\
 \hline  Familon & $1.4\cdot 10^{-13}$ &$1\cdot10^{-49}$&$3\cdot10^{-4}$\\
 \hline  Majoron & $3.4\cdot 10^{-14}$ &$5\cdot 10^{-51}$ & $1.5\cdot
 10^{-5}$\\
 \hline \end{tabular} \end{center}

 \section{The registration}

 To see the obtained LGP the special simple detector is proposed. It should be
some pin--type lead rod with radius about 2 cm and length about 100 m, placed
in vacuum behind a shield to get rid of the background (Fig.2). The round
scinillator with diameter in 1-3 m in the end of this device should detect
particles produced in lead.

\begin{figure}
\centering\includegraphics[bb=170 160 430 650, scale=0.5]{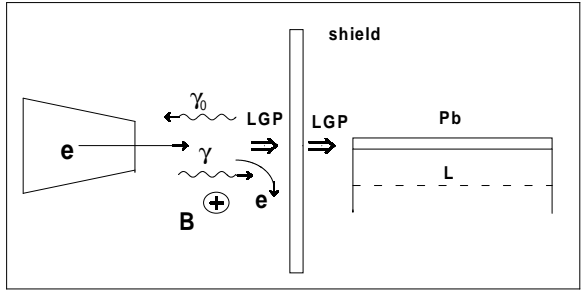}
\caption{The scheme of LGP production and registration.}
\end{figure}

The LGP should interact with lead nuclei and production of hadrons:
\begin{equation}
    X + Pb \to h \quad {\rm (hadrons) }
\label{18}
\end{equation}
    This reaction will be observed as the production of hadron jets with
 total energy $\sim \epsilon _X$ and characteristic transverse momentum
 $p_\perp \sim 300$ MeV/c.
    Let us evaluate the number of events for the case of  standard axion.
         The cross-section of reaction (\ref{18}) is $\sim A$ times as large
 as the cross-section of the axion--nucleon interaction, $\sigma _{an}$,
 where $A$ is the number of nucleons in a nucleus.
         The cross-section  $\sigma _{an} \approx
  f_{a\pi }\sigma _{\pi n}(v/f_{pq})^2$, where $f_{a\pi }$ is the amplitude
 of the axion--pion transition for the standard axion, $f_{a\pi }=
 2\cdot 10^{-4}$\cite{SW}.
 Therefore\footnote{The cross-section of lepton pairs production in
Bete-Haitler
 reaction $X+Pb \to e^+e^-+ \cdots\;, X+Pb \to \mu ^+\mu ^- +\cdots $
 is approximately an order less then cross-section discussed.}
 \begin{equation}
 \sigma ( {X_a+Pb \to h}) \approx Af_{a\pi}^2(v/f_{pq})^2\sigma _{\pi n}
 \approx 5\cdot 10^{-34}cm^{2}.
\end{equation}
  With this cross section on the path of lead of 100 m
  one event (\ref{3}), (\ref{18}) per hour will be observed.
  The increase of additional $U(1)$ symmetry scale to one order
  gives the decrease of the events number in lead of four orders,
 because we really think it is possible to reach $f_{pq} \sim 10$ TeV in
 this experiment.

         The background for the reaction (\ref{3}), (\ref{18}) will be
produced via
    the high energy photon interaction with the matter of detector. With
 good wall behind vacuum camera our lead
detector can destinate only neutrinos produced in this wall.
Their angular spread ($>300\rm{ MeV}/E$) is much higher that of LGP's
calculated above. Besides, energies of these neutrino should be less than the
 energies of LGP's.

\section{Acknowledgments}

I would like to thank I.F.Ginzburg for improving
English in the manuscript and valuable remarks.
 The author thanks R.N.Mohapatra
for the criticism of majoron problem.


\begin{thebibliography}{99}
\bibitem{Ans}
{\it A.A.Anselm, N.G.Uraltsev, V.A.Khoze}, Uspekhi Fiz. Nauk ,
{\bf 145}, 1, (1985)
\bibitem{Sik}
{\it P.Sikivie} Axions Searches, hep-ph/0002154
 \bibitem{GKST}
 {\it I.Ginzburg, G.Kotkin, V.Serbo, V.Telnov,} Pizma ZhETF,
 {\bf 4}, 514, (1981);
 Nucl.  Instr.  and  Methods (NIM), {\bf 205}, 47, (1983);
 {\it I.F. Ginzburg, G.L.  Kotkin,  S.L.  Panfil, V.G. Serbo, V.I. Telnov,}
 Nucl.  Instr.  and  Methods, {\bf 219}, 5,(1984).
\bibitem{Gin}
{\it I.F.Ginzburg, } Nucl.Instr. and Methods (NIM), {\bf A 355} 63,(1995)
 \bibitem{Tel}
 {\it V.I.Telnov, } Nucl.Instr. and  Methods, {\bf 24}, 72, (1990);
 Talk at ICFA Workshop {\it Quantum Aspect of Beam
 Physics}, Monterey, CA, USA, January 4-9,(1998); hep-ex/9805002.
\bibitem{Brink}
Conceptual Design of a 500 GeV $e^+e^-$ Linear
Collider with Integrated X-ray Laser Facility  Ed. {\it R.Brinkmann et al.}
 DESY 1997-048, ECFA-1997-182.
\bibitem{Pol1}
{\it S.I.Polityko,} Yad. Fiz. {\bf 43} 144, (1986) [Sov. J. Nucl. Phys.
{\bf 43}, 93 (1986)]
\bibitem{Brod}
{\it S.Brodsky, E.Mottola, I.Muzinich, M.Soldate}, Phys.Rev.Lett.
{\bf 56} 1763,(1986).
\bibitem{Pol2}
{\it S.I.Polityko,} Yad. Fiz. {\bf 56} 146,(1993)
[Phys. At. Nucl. {\bf 56}, 83,(1993)]
\bibitem{GKP}
{\it I.F.Ginzburg, G.L.Kotkin, S.I.Polityko, } Yad. Fiz.,{\bf 56}, 65,(1993)
[Phys. At. Nucl. {\bf 56}, 1487,(1993)]
\bibitem{SW}
{\it S.Weinberg}, Phys.Rev.Lett. {\bf 40} 223,(1978).
\bibitem{FW}
{\it F.Wilczek,} Phys.Rev.Lett. {\bf 40} 279,(1978).
\bibitem{GGR}
{\it G.G.Raffelt,}  Phys.Rep.{\bf 198} (1990).
\bibitem{Z}
{\it A.R. Zhitnitsky,} Yad.Fiz. {\bf 31}, 497,(1980) [Sov. J. Nucl. Phys.
{\bf 31}, 260, (1980)]
\bibitem{DFS}
{\it M.Dine, W.Fischler, M.Srednicki}, Phys.Lett. {\bf 104B},199, (1981).
\bibitem{K}
{\it J.E.Kim,} Phys.Rev.Lett. {\bf 43}, 103 (1979).
\bibitem{SVZ}
{\it M.A.Shifman, A.I.Vainstein, V.I.Zakharov,} Nucl.Phys.
{\bf B166} 493,(1980).
\bibitem{Ber}
{\it Z.G.Berezhiani, A.S.Sakharov, M.Y.Khlopov}, Yad. Fiz.
{\bf 55} 1918,(1992).
\bibitem{Kim}
{\it J.E.Kim}, preprints SNUTP 98-141, KIAS-P98047, astro-ph/9812257
\bibitem{PDG}
{\it Particle Date Group}, Phys.Rev. {\bf D54},1, (1996).
\bibitem{GR}
{\it G.B.Gelmini, M.Roncadelli,} Phys.Lett.{\bf 99B}, 401, (1981).
\bibitem{RM}
{\it Y.Chikashige, R.N.Mohapatra, R.D.Peccei,} Phys.Lett.{\bf 98B}, 265, (1981).
\bibitem{Cheng}
{\it H.Y.Cheng,} Phys.Rev. D, {\bf 36}, no.6, 1649, (1987).
\bibitem{Ritus}
{\it V.I.Ritus,} Trudy FIAN, {\bf 111} ,(1979) [in Russian].
\bibitem{LD}
{\it V.Berestetskii, E.Lifshitz and L.Pitaevskii,} Quantum Electrodynamics,
Pergamon press, Oxford, (1982).
\end{thebibliography}
\end{document}